\documentstyle[aps,preprint,eqsecnum]{revtex}
\begin{document}
\draft
\preprint{Aug. 5, 1995}
\title{
Quantum Group, Bethe Ansatz \\
and \\
Bloch Electrons in a Magnetic Field
}
\author{Yasuhiro Hatsugai$^{1*}$, Mahito Kohmoto$^2$,
and Yong-Shi Wu$^3$}
\address{
$^1$ Department of Applied Physics, 
 University of Tokyo,
 7-3-1 Hongo, Bunkyo-ku, Tokyo 113, Japan. \\
$^2$ Institute for Solid State Physics,
 University of Tokyo,
 7-22-1 Roppongi Minato-ku, Tokyo 106, Japan. \\
$^3$ Department of Physics, University of Utah,
 Salt Lake City, Utah 84112, U.S.A.
}

\maketitle
\begin{abstract}
The wave functions for two dimensional  Bloch electrons in a uniform
magnetic field at the mid-band points are studied
with the help of the algebraic structure of
the quantum group $U_q(sl_2)$. A linear combination of its
generators gives the Hamiltonian. We obtain analytical and 
numerical solutions for the wave functions by solving 
the Bethe Ansatz equations, proposed by Wiegmann and Zabrodin 
on the basis of above observation. 
The semi-classical case with the flux per plaquette 
$\phi=1/Q$ is analyzed in detail, by exploring a structure 
of the Bethe Ansatz equations.
We also reveal the multifractal structure of the 
Bethe Ansatz solutions and corresponding wave functions 
when $\phi$ is irrational, such as the golden or silver mean. 

\end{abstract}

\narrowtext
\section{Introduction}
The effects of a magnetic field on electrons in quantum
 regime is a fascinating problem with very rich physics. 
The path-dependent geometric phase of the wave function 
is known to play a fundamental role in the problem. 
\cite{geom} For example, the geometrical 
phase and other geometrical considerations, 
such as the gauge invariance \cite{lgh} and 
the effect of edges \cite{halp}, are very crucial 
to the quantum Hall effect. \cite{qh,geom}  Moreover, the 
quantized Hall conductance has a topological character,
both for a system without boundary (periodic boundary 
conditions) \cite{tknn} and for a system with
boundaries (edges) \cite{edge_yh}. Thus geometry and 
topology are of central importance in this realistic 
condensed-matter problem. 

In this paper we will show that a certain exotic 
algebraic structure plays an important role as well, 
if one includes the effects of a periodic potential 
in this problem. This is not too surprising, since there
are two fundamental periods in the problem: One is the
original period in the geometric phase, and the other
is that of the periodic potential. Interplay of the two
intrinsic periods brings more and interesting structures 
and, therefore, much richer physics into the problem. 

Normally a system with strong periodic potential is 
modeled by the tight-binding Hamiltonian and the effects  
of the magnetic field is included by the Pierls
substitution. This procedure is usually motivated as 
an approximation. However, the gauge invariance 
is preserved by the substitution, i.e.\  
both of the original continuum and the lattice 
systems have the same local $U(1)$ symmetry. 
Thus in our opinions, the Pierls substitution 
has captured a fundamental feature of the problem and,
therefore, has a meaning more fundamental than as merely
an approximation. 

So in the problem of a Bloch electron in a uniform
magnetic field, one can use the flux per plaquette,
$\phi$, to characterized the system. The equations of 
motion can then be reduced to one-dimensional ones,
which appear in many different physical contexts, 
ranging from the quantum Hall effect\cite{tknn,edge_yh} 
to quasi-periodic systems \cite{hk}. 
 When $\phi$ is irrational, 
that is, the magnetic flux is incommensurate with 
the periodic potential, the spectrum is known to 
have an extremely rich structure like the Cantor 
set and to exhibit a multifractal behavior.\cite{ht,hk}
Another interesting case is the weak field limit. 
When the flux is small, the semi-classical treatment 
of the WKB type is justified. 

Recently a relationship is found between this physical
problem and a mathematically new concept, the 
so-called quantum group.\cite{wz} The essential point
is that at least for some special points in the 
spectrum, e.g.\ at the mid-band points, the Hamiltonian
(in certain gauge) can written as a linear combination 
of generators of a quantum group, called $U_q(sl_2)$. 
On the basis of this observation, Wiegmann and Zabrodin
has proposed Bethe-Ansatz type equations for 
corresponding wave functions. This opened the 
possibility of solving the problem analytically. 
Indeed, we have turned the dream into reality:
We have found explicit analytical solutions to this 
problem at the first time in the literature. \cite{hkw} 
The algebraic structure of $U_q(sl_2)$ and the 
associated Bethe Ansatz equations also gives us a new
method to solve the problem numerically. Especially
when we study the irrational limit of a well-organized 
sequence of rational fluxes, the new method is 
convenient and beneficial in revealing the multifractal
behavior, as we briefly reported also in ref. \cite{hkw}.

In this way, it becomes clear that in addition to the 
underlying geometrical and topological structures, 
the problem of a Bloch electron in magnetic field 
also has an interesting, fundamental algebraic structure 
hidden in it. Previously there were works in the literature 
about possible relation of the quantum groups to
physics. Many of the works, however, deal with a sort of
artificial (rather realistic) $q$-deformation of an 
original physical problem. On the other hand, for the 
Bloch electrons, the quantum group is for a realistic 
physical problem: We are dealing with ordinary electrons, 
not the {\it q}-electrons. Therefore we have shown that
quantum groups can have a fundamental meaning for
observable real phenomena. 

In this paper we are going to report on more details of
our published results and to present a number of new 
solutions and results. We will describe our analytical 
solutions in detail and in a self-contained manner. 
We also present more numerical solutions of the Bethe 
Ansatz equations when the analytical ones are not available. 
In the next section, Sec. \ref{sec:bloch}, we review 
the Bloch electron and the magnetic translations. 
In Sec. \ref{sec:qgroup} and  \ref{sec:func_ba}, we 
summarize necessary stuff about the quantum group 
$U_q(sl_2)$, its representations and the functional 
Bethe Ansatz equations following Wiegmann and Zabrodin.
\cite{wz} Then in Sec. \ref{sec:exp}, we explain in detail
our way of analytically solving the Bethe Ansatz equations
at a particular mid-band point $E=0$.
Sec. \ref{sec:exp_wfn} is devoted to the analysis of
the wavefunction by using the explicit solutions obtained
in the previous Sec. \ref{sec:exp}. We also present 
a multifractal analysis of the explicit wavefunctions
for irrational fluxes, such as the golden and silver means.
Finally in Sec. \ref{sec:num}, we present numerical 
solutions to the Bethe Ansatz equations at other mid-band
points than $E=0$. Some technical details are put into several
Appendices for the convenience of reader.

\section{Bloch Electrons and The Magnetic Translations}
\label{sec:bloch}
A tight-binding electron in a magnetic field 
is described by  the Hamiltonian
\begin{equation}
  \label{q:hmlt_by_T}
H    =   T_x + T_y +T_{x}^\dagger+T_{y}^\dagger, \\
\end{equation}
where $T_x$ and $T_y$ are the {\it covariant translation} 
operators defined by
\begin{eqnarray}
T_x & = &  \sum_{m,n}
   c_{m+1,n}^\dagger  c_{m,n} e^{ i \theta^x_{m,n}}, \\
T_y & = &  \sum_{m,n}
   c_{m,n+1}^\dagger  c_{m,n} e^{ i \theta^y_{m,n}} ,
  \label{q:cov_T}
\end{eqnarray}
where  $c_{m,n}$ is the annihilation operator for an electron
at site $(m,n)$.
The phase factors
$ \theta^x_{m,n}$ and
$ \theta^y_{m,n}$ are related to a flux per plaquette $\phi_{m,n}$ 
at $(m,n)$ by
\begin{equation}
  \label{eq:lattice_rotation}  
{\rm rot}_{(m,n)}\theta  =
 \Delta_x \theta_{m,n}^y- \Delta_y \theta_{m,n}^x = 2 \pi \phi_{m,n},
\end{equation}
where the difference operators $\Delta_x$ and
 $\Delta_y$ operate on a lattice 
function $f_{m,n}$ as 
\begin{eqnarray}
  \label{eq:diff}
  \Delta_x f_{m,n} & = &f_{m+1,n}-f_{m,n}\\
  \Delta_y f_{m,n} & = &f_{m,n+1}-f_{m,n}.
\end{eqnarray}
As mentioned in the introduction, 
$H$ has a local $U(1)$ gauge symmetry, 
{\it i.e.}, it is invariant under 
\begin{eqnarray}
\label{eq;gauge}
c_i&\to&\Omega_i c_i \\
e^{i \theta_{ij}} &\to& \Omega_i e^{i \theta_{ij}}
\Omega_j^{-1} \\
| \Omega_j | & = & 1,\ \ \forall j=(m,n).
\end{eqnarray}

The covariant translations $ T_x$  and $ T_y$ 
do not commute with each other. For a one-particle state 
$ |\Psi_{m,n}\rangle = c_{m,n}^\dagger |0\rangle$
which is localized at the site $(m,n)$,
their operation satisfies
\begin{equation}
T_y T_x|\Psi_{m,n}\rangle=e^{i 2 \pi \phi_{m,n}} 
T_x T_y|\Psi_{m,n}\rangle.
  \label{eq:non_commute}
\end{equation}

Now let us assume that the magnetic field is uniform,
so that we can choose a gauge in which both 
$\Delta_x \theta_{m,n}^y$ and $\Delta_y \theta_{m,n}^x$
are constants independent of the site $(m,n)$. 
In this case, there is an important symmetry of the 
Hamiltonian, described by the magnetic translation
operators  $\hat T_x$  and $\hat  T_y$ which 
commute with the Hamiltonian
\begin{equation}
[H, \hat  T_x] =[H, \hat  T_y] =0.
  \label{eq:commutation_relation}
\end{equation}
They are explicitly given by
\begin{eqnarray}
\hat T_x & = &  \sum_{m,n}
   c_{m+1,n}^\dagger  c_{m,n} e^{  i \chi^x_{m,n}}, \\
\hat T_y & = &  \sum_{m,n}
   c_{m,n+1}^\dagger  c_{m,n} e^{  i \chi^y_{m,n}} ,
  \label{q:magnetic_translation}
\end{eqnarray}
where the phases $\chi_{m,n}^{x,y}$ satisfy the relations
\begin{eqnarray}
  \label{eq:chi}
  \Delta_x \chi_{m,n}^x &=& \Delta_x\theta_{m,n}^x ,\\
  \Delta_y \chi_{m,n}^x &=& \Delta_x\theta_{m,n}^y 
\ \ (=\Delta_y \theta_{m,n}^x+2 \pi \phi_{m,n}),\\
  \Delta_x \chi_{m,n}^y &=& \Delta_y\theta_{m,n}^x
\ \ (=\Delta_x \theta_{m,n}^y-2 \pi \phi_{m,n}) ,\\
  \Delta_y \chi_{m,n}^y &=& \Delta_y\theta_{m,n}^y. 
\end{eqnarray}
In the parenthesis, we have used Eq.~(\ref{eq:lattice_rotation}).
We can easily solve them as 
\begin{eqnarray}
  \label{eq:chi_expression}
  \chi_{m,n}^x&=&\theta_{m,n}^x
+2 \pi n \phi_{m,n} \\
  \chi_{m,n}^y&=&\theta_{m,n}^y
-2 \pi m \phi_{m,n},                  
\end{eqnarray}
  which are gauge dependent. 

In the following, let us assume the magnetic field is uniform         
and rational, that is, $\phi_{m,n}=\phi=P /Q$ with mutually
prime integers $P$ and $Q$.
We can take a ``diagonal'' gauge\cite{kh,wz} as
\begin{equation}
  \label{eq:diagonal}
  \theta_{m,n}^x=\pi\phi(m+n),\ \ 
  \theta_{m,n}^y=-\pi\phi(m+n+1),
\end{equation}
which is derived from the Landau gauge in $(1,1)$ direction.
Correspondingly, we define  magnetic translations along 
$(1,1)$ and $(-1,1)$ directions $\hat T_+$ and $ \hat T_-$ 
respectively as 
\begin{eqnarray}
  \label{eq:diag_trans}
  \hat T_+ &=& \hat T_x \hat T_y , \\
  \hat T_- &=& \hat T_x ^{-1} \hat T_y .
\end{eqnarray}
They commute with the Hamiltonian, 
$[H,\hat T_+]=[H,\hat T_-]=0$.
They do not commute with each
other.   
However, one can explicitly check that $\hat T_+^Q$ 
and $\hat T_-$ commute with each other. Thus we can 
take simultaneous eigenstates  of $H$ and $\hat T_{+}^{Q}$ 
and $\hat T_{-}$, which are specified by the 
momentum in the magnetic Brillouin zone  
(the Bloch theorem) as
\begin{eqnarray}
  \label{eq:mom_def1}
  \hat T_+^Q |\Phi(p_+,p_-)\rangle & = &
 e^{i 2 Q  p_+} |\Phi(p_+,p_-)\rangle , \\
  \label{eq:mom_def2}
  \hat T_- |\Phi(p_+,p_-)\rangle & = & 
e^{i   p_-} |\Phi(p_+,p_-)\rangle, 
\end{eqnarray}
where $|\Phi\rangle $ is a one particle state
spanned as 
\begin{equation}
  \label{eq:one_p_state}
|\Phi(p_+,p_-)\rangle  =
 \sum_{m,n} \Psi_{m,n}(p_+,p_-) c_{m,n}^\dagger |0\rangle.
\end{equation}
  From Eqs. ~(\ref{eq:mom_def1}) and (\ref{eq:mom_def2}, we have, 
\begin{eqnarray}
    \label{eq:bloch}
     \Psi_{m,n}(\mbox{\boldmath $p$}) &=& 
e^{ip_-(m-n)+ip_+(m+m)} \psi_{m+n}(\mbox{\boldmath $p$}) \\
 \psi_{k+2Q}(\mbox{\boldmath $p$})& = & 
 \psi_{k}(\mbox{\boldmath $p$}), \ \ k=1,\cdots,2Q.
\end{eqnarray}
The Schr\"odinger equation, 
$H |\Phi(\mbox{\boldmath
$p$})\rangle =  E(\mbox{\boldmath $p$}) |\Phi(\mbox{\boldmath
$p$})\rangle$
 is written by
$\psi_k(\mbox{\boldmath $p$})$ as
\begin{eqnarray}
  \label{eq:sh_eq}
 ( e^{i(p_++p_-)} q^{-k} +  e^{i(p_+-p_-)} q^{k+1} ) 
\psi_{k+1}(\mbox{\boldmath $p$})
&+&\nonumber \\
 ( e^{-i(p_++p_-)} q^{k-1} &+&  e^{-i(p_+-p_-)} q^{-k} )
 \psi_{k+1}(\mbox{\boldmath $p$})
= 
\psi_k(\mbox{\boldmath $p$}),
\end{eqnarray}
where $q$ is defined as 
\begin{equation}
  \label{eq;def_q}
  q=e^{i\pi\phi}=e^{i\pi{\frac P Q}}.
\end{equation}

Using base 
$\mbox{\boldmath$\psi$}(\mbox{\boldmath$p$})  = 
[ \psi_1(\mbox{\boldmath$p$}),
\cdots,\psi_{2Q}(\mbox{\boldmath$p$})]^t$,
Eq.~(\ref{eq:sh_eq}) is 
written in the matrix form as 
$
H(\mbox{\boldmath$p$})
 \mbox{\boldmath$\psi$}(\mbox{\boldmath$p$})  = 
E(\mbox{\boldmath$p$}) \mbox{\boldmath$ \psi$}
(\mbox{\boldmath$p$}),
$ 
with
\begin{eqnarray}
H(\mbox{\boldmath$p$}) & = & 
e^{i(p_++p_-)} Y^{-1} X^{-1} + e^{i(p_+-p_-)} X^{-1} Y \nonumber \\
& +  &  e^{-i(p_++p_-)}  X Y  + e^{-i(p_+-p_-)} Y^{-1} X ,\nonumber\\
& = & ( e^{-i(p_++p_-)}  X +e^{i(p_+-p_-)} X^{-1}) Y \nonumber \\
& +&  Y^{-1} ( e^{-i(p_+-p_-)} X + e^{i(p_++p_-)}  X^{-1}  )
  \label{eq:matrix_form}
\end{eqnarray}
where $2Q\times 2Q$ matrices $X$ and $Y$ are given by
\begin{eqnarray}
X & = & \left[
\begin{array}{ccccc}
 \ 0 \ \ & \ \ \ \   &\ \ \   &\ \ \  &\ \  1\ \\
\ 1 \ & 0 &   &   &  \\
 0 & 1 & 0 &   &  \\
   &   & \ddots & \ddots &  \\
   &   &  & 1 & 0 
\end{array}
\right], Y = 
{\rm diag} (q^1,q^2,\cdots,q^{2Q}).
  \label{eq:cyclic_X}
\end{eqnarray}
They satisfy  relations
\begin{eqnarray}
X^{2Q}&=&Y^{2Q}=I_{2Q}, 
  \label{eq:cyclic_relation} \\
qXY& =& YX, \ X^T=X^{-1}, Y^T=Y
  \label{eq:comm_relation}
\end{eqnarray}
The group generated by $X$ and $Y$ with the relations above
is the Heisenberg-Weyl group. 

The solvability of the problem of the Bloch electrons
by the Bethe Ansatz lies in the fact that
the Hamiltonian $H(\mbox{\boldmath$p$})$ 
has a higher symmetry than the Heisenberg-Weyl group
at special momentum points. \cite{wz}
Actually 
$H(\mbox{\boldmath$p$})$ 
is written  as 
\begin{equation}
H_{MB}(p_-)\equiv H(p_+=\pi/2,p_-)=i(q-q^{-1})
( 
e^{-ip_-} \rho_C(B) + 
e^{ip_-} \rho_C(C)),
\label{eq:def_midband}
\end{equation}
at special momentum lines $p_+=\pi/2 \bmod{\pi/Q}$, 
usually called as the mid-band lines. ($p_-$ is a 
free parameter, but later we will see that the
energy is independent of it.). Here
$\rho_C:U_q(sl_2)\to M_{2Q}$ ($2Q$-dimensional matrices) 
is a cyclic representation of the  quantum group
$U_q(sl_2)=\{A,B,C,D\}$. 
It is explicitly given by 
\begin{eqnarray}
\label{eq:cyclic_rep_A}
\rho_C(A)& = & q^{-(Q-1)/2} X^{-1}, \\
\label{eq:cyclic_rep_B}
\rho_C(B)& = & -(q-q^{-1})^{-1}(X -
X^{-1})Y, \\
\label{eq:cyclic_rep_C}
\rho_C(C)& = & -(q-q^{-1})^{-1}Y^{-1}(X
-  X^{-1}), \\
\label{eq:cyclic_rep_D}
\rho_C(D)& = & q^{ (Q-1)/2} X. 
\end{eqnarray}
The representation $\rho_C$ will be used to reveal
the quantum group structure of the problem 
in the next section.

\section{the quantum group and its representation}
\label{sec:qgroup}
Historically quantum groups arise from mathematical 
structures appearing in the exactly solvable 
one-dimensional quantum many-body models. 
However, the simplest case, $U_q(sl_2)$, can be
understood as an {\it a priori} deformation 
of the ordinary Lie algebra $sl_2$:
$U_q(sl_2)$ is obtained by the $q$-deformation 
(i.e.\  an appropriate ``extension'' with a new 
complex parameter $q$, which preserves many good 
properties) of $sl_2$ and, conversely, $sl_2$ 
is recovered from $U_q(sl_2)$ by taking a 
classical limit $q\to 1$.

More precisely, $U_q(sl_2)$ is an algebra
generated by four generators $\{  A, B, C, D \}$
with multiplication and addition (with complex
coefficients), subject to the following 
defining relations
\begin{eqnarray}
AD &=& DA=1, \\
AB &=& qBA,\ \ BD=qDB, \\
DC &=& qCD,\ \ CA=qAC, \\
 {[ B,C ]}  &=& {\frac{A^2-D^2}{q-q^{-1}}}
\label{eq:defing_relation}
\end{eqnarray}
This is the $q$-deformation of 
the Lie algebra $sl_2=\{ S_+, S_-, S_3\}$,
which is defined by the relations 
$[S_+,S_-]=2S_3,\ [S_3,S_{\pm}]=\pm S_{\pm}$. 

It is well known that $sl_2$ has a $(2j+1)$ dimensional
(spin-$j$) representation which is realized 
by the differential operators 
\begin{eqnarray}
\label{eq:sl2}
\rho_h(S_3)&=&z\partial_z-j, \\
\rho_h(S_+)&=&z(2j-z \partial_z), \\
\rho_h(S_-)&=&\partial_z.
\end{eqnarray}
acting on  a function $\Psi(z)$.
This finite-dimensional representation has both a 
highest weight and a lowest weight, since 
$\rho_h(S_+) z^{2j} = \rho_h(S_-) z^0=0$;
thus, it is a $2j+1$-dimensional representation 
with $\Psi(z)$ a polynomial of degree $2j$. 
Correspondingly, $U_q(sl_2)$ also has a highest-weight 
representation, realized by the $q$-difference operators:
\begin{eqnarray}
\rho_h(A)\Psi(z)& = & q^{-j} \Psi(qz), \\
\label{eq:high_rep_A}
\rho_h(B)\Psi(z)& = &
 (q-q^{-1})^{-1}z[q^{2j}\Psi(q^{-1}z)-q^{-2j}\Psi(qz)], \\
\label{eq:high_rep_B}
\rho_h(C)\Psi(z)& = &
 -(q-q^{-1})^{-1}z^{-1}[\Psi(q^{-1}z)-\Psi(qz)], \\
\label{eq:high_rep_C}
\rho_h(D)\Psi(z)& = & q^{j} \Psi(q^{-1}z). 
\label{eq:high_rep_D}
\end{eqnarray}
It has also a highest weight and a lowest weight:
$\rho_h(B) z^{2j} = \rho_h(C) z^0=0$. 
In the standard bases, $\Psi(z)=z^n$, ( $n=0,1,\cdots,Q-1$),
they operate as
$\rho_h(A) z^n=q^{n-j}z^n$,
$\rho_h(D) z^n=q^{j-n}z^n$,
$\rho_h(B) z^n=[2j-n]_qz^{n+1}$, and
$\rho_h(C) z^n=[n]_qz^{n-1}$ where $[n]$ is 
a $q$-integer defined as 
\begin{equation}
  \label{eq:qint}
  [n]_q={\frac {q^n-q^{-n}}{q-q^{-1}}},
\end{equation}
which approaches to $n$ in the limit $q \to 1$.
Thus they can be represented as matrices, given by 
\begin{eqnarray}
\rho_h(A)&:& \ {\rm diag } (q^{-j},q^{1-j},\cdots,q^{j-1},q^{j}) 
  \label{eq:highest_A}\\
\rho_h(D)&:& \ { \rm diag}
(q^{j},q^{-1+j},\cdots,q^{-j+1},q^{-j}) 
  \label{eq:highest_D}\\
\rho_h(B)&:& \ \left[
\begin{array}{ccccc}
 \ 0  \ & \    &\    &\   & 0  \\
\ [2j]_q \ & 0 &   &   &  \\
 0 & [2j-1]_q & 0 &   &  \\
   &   & \ddots & \ddots &  \\
0   &   &  & [1]_q & \ 0 
\end{array}
\right]
  \label{eq:highest_B}
\\
\rho_h(C)&:& \ \left[
\begin{array}{ccccc}
 \ 0 \ \ & [1]_q  &\ \ \   &\ \ \  & 0  \\
\  \ & 0 & [2]_q  &   &  \\
   &   & \ddots & \ddots &  \\
  &  &  &0   &\  [2j]_q \  \\
0   &   &  &  & 0 
\end{array}
\right]
  \label{eq:highest_C}
\end{eqnarray} 
Notice that the dimension of the matrix above is 
$Q\equiv 2j+1$ and the matrices are all tri-diagonal.

Up to this point, $U_q(sl_2)$ is quite analogous 
to the $sl_2$. However, when $q$ is a root of unity, 
there is another representation, called as a cyclic 
representation, whose existence is characteristic to 
$U_q(sl_2)$. ($sl_{2}$ does not have cyclic representations.)
Our case, $q^{2Q}=1$, is such a case.  Here a 
$2Q$-dimensional cyclic representation is derived from 
the highest-weight representation of 
spin-$j=(Q-1)/2$ as follows. First define a cyclic basis 
$\{\Psi_n,\ n=1,\cdots, 2Q \ \pmod{2Q} \}$ for the
cyclic representation (satisfying $\Psi_{n+2Q}=\Psi_n$) 
by
\begin{equation}
\Psi_n\equiv \Psi(q^n),
\label{eq:define_base}
\end{equation}
where $\Psi(z)$ is the polynomial in the spin-$j=(Q-1)/2$ 
highest-weight representation.
In this realization, operations of the cyclic 
representation are explicitly given by
\begin{eqnarray}
\rho_C(A)\Psi_n& = & q^{-(Q-1)/2} \Psi_{n+1}, \\
\label{eq:base_cyclic_rep_A}
\rho_C(B)\Psi_n& = & 
-(q-q^{-1})^{-1}
(q^{n+1}\Psi_{n+1}-q^{n-1}\Psi_{n-1}), \\
\label{eq:base_cyclic_rep_B}
\rho_C(C)\Psi_n& = & (q-q^{-1})^{-1}
(q^{-n}\Psi_{n+1}-q^{-n}\Psi_{n-1}), \\
\label{eq:base_cyclic_rep_C}
\rho_C(D)\Psi_n& = & q^{ (Q-1)/2} \Psi_{n-1}. 
\label{eq:base_cyclic_rep_D}
\end{eqnarray}
If we write them in matrix form, we get
Eqs.~(\ref{eq:cyclic_rep_A})-~(\ref{eq:cyclic_rep_D}). 
We note that $\rho_{C}(B)\Psi_{2Q}$ returns to a linear
combination of $\Psi_{1}$ and $\Psi_{2Q-1}$. So there
is no highest-weight state (or vector) $\Psi$ that 
can be annihilated by $\rho_{C}(B)$: $\rho_{C}(B)\Psi=0$.
Similar things happen to $\rho_{C}(C)$: There is no
lowest-weight state annihilated by $\rho_{C}(C)$.
These are the reasons why this representation is
called cyclic. 

Thus, the highest-weight representation of dimension $Q$ 
is related to the cyclic representation of dimension $2Q$.
This establishes the following results about the 
spectrum and the wavefunction of the Bloch electron
at the mid-band momenta. The spectrum is given by the 
eigenvalues of the following $Q\times Q$  
tri-diagonal hermite matrix.
\begin{equation}
  \label{eq:tri-hmlt}
  H_{MB}^{tri}(p_-) = i(q-q^{-1})
 \left[
\begin{array}{cccccc}
 0  & [1]_q e^{-ip_-} &  & & &   \\
 {[1]}_q e^{+ip_-}  & 0 & [2]_q e^{-ip_-} & & O  &  \\
    & [2]_q e^{+ip_-}  & 0 & \ddots & &  \\
    &   & \ddots & \ddots\hfil   & [Q-2]_q e^{-ip_-}  &    \\
    & O  &  &   [Q-2]_q e^{+ip_-}& 0   &  [Q-1]_q e^{-ip_-}  \\
    &   &  &  & [Q-1]_q e^{+ip_-}  &0 
\end{array}
\right],
\end{equation}
where we have used the fact $[Q-j]_q=[j]_q$.
When we take large $Q$, 
the eigenvalues of the tri-diagonal matrix gives the
so-called  Hofstadter's butterfly. 
In Figs. \ref{f:ene}, we show eigen values of the 
$H_{MB}^{tri}$ as a function of $\phi=\frac 1 \pi \Im  \log q$ for
$Q=23$ and $Q=401$. 
Since the eigen values of the tri-diagonal
matrix  does not depend on the phase of the off-diagonal matrix
elements, the energy  does not depend on $p_-$. 
The wavefunction of the Bloch electron
at the mid band momentum of the $j$-th band is obtained from
the $j$-th eigen vector $\{v^j_m\},~ (m=0,1,\cdots,Q-1)$
as
\begin{equation}
  \label{eq:wfn_general}
  \Psi^j_n = \Psi(q^n)=\sum_{m=0}^{Q-1} v^j_m q^{nm}.
\end{equation}
The dimension of the original Hamiltonian $H_{MB}$ is $2Q$
and the tri-diagonal functional Hamiltonian is $H_{MB}^{tri}$ is
$Q$. Actually,  $H_{MB}$ has doubly degenerate eigen values.  
The above state eq.~(\ref{eq:wfn_general})
is degenerate with the other state
\begin{equation}
  \label{eq:wfn_general_o}
 \Psi'^j_n = \Psi(q^{-n})=\sum_{m=0}^{Q-1} v^j_m q^{-nm}.
\end{equation}
These two states have the same energy and they are orthogonal
each other as 
\begin{equation}
  \label{eq:orth}
  \sum_{n=0}^{2Q-1}\Psi^j_n   \Psi'^{j*}_n
= \sum_{0\le m,m'\le Q-1} v_m v_{m'}^* \frac{1-q^{2Q(m+m')}}
{1-q^{(m+m')}} = 0
\end{equation}
Thus we can obtain all solutions of the original
problem from the functional equation.

\section{the functional bethe ansatz for the bloch electrons}
\label{sec:func_ba}
In this section, we derive the functional Bethe Ansatz
equation following Wiegmann and Zabrodin \cite{wz}. 
We denote $p_-$ by $p$, in the following.
The  Schr\"{o}dinger equation 
at a mid band is 
\begin{equation}
H_{MB}(p)\psi_n=i(q-q^{-1})
[ 
e^{-ip} \rho_C(B) + 
e^{ip} \rho_C(C)]\Psi_n = E \Psi_n. 
\label{eq:sh_eq_c}
\end{equation}
Recalling that the cyclic representation is derived from the
highest weight representation, let us define the functional 
equation 
\begin{eqnarray}
i(q-q^{-1})
 [ 
e^{-i p} \rho_h(B) + 
e^{i p} \rho_h(C)]\Psi(z) & = & \nonumber \\
i(e^{-i p}qz+e^{i p}z^{-1})\Psi(qz)
- -i(q^{-1}e^{-i p}z+e^{i p}z^{-1})\Psi(q^{-1}z) & = & E \Psi(z).
\label{eq:sh_eq_h}
\end{eqnarray}
Once we have a solution of the functional equation 
Eq.~(\ref{eq:sh_eq_h}), 
we can recover all the solutions of Eq.~(\ref{eq:sh_eq_c})
using Eq.~(\ref{eq:wfn_general_o}) and Eq.~(\ref{eq:wfn_general}).

Since $\Psi(z)$ is a polynomial of finite degree, 
it can be factorized as
\begin{equation}
\Psi(z)=\prod_{m=1}^{Q-1}(z-z_m(p)).
\label{eq:factorized}
\end{equation}
First let us absorb the ${p}$ dependence of
Eq.~(\ref{eq:sh_eq_h})  by the gauge transformation
\begin{eqnarray}
\bar z&=& z e^{-i p}, \\
\label{eq:tra_z}
\bar z_m&=& z_m(p) e^{-i p}, \\
\label{eq:tra_zm}
\bar\Psi(\bar z)& = &\prod_{m=1}^{Q-1}(z-z_m(p)) 
\nonumber \\
       & = & e^{i(Q-1)p}\prod_{m=1}^{Q-1}(\bar z-\bar z_m ).
\label{eq:tra_psi}
\end{eqnarray}
Thus we have
\begin{equation}
i(q\bar z+\bar z^{-1})\bar\Psi(q\bar z)
- -i(q^{-1}\bar z+\bar z^{-1})\bar \Psi(q^{-1}\bar z)  = 
 E \bar \Psi(\bar z).
\label{eq:sh_eq_f_b}
\end{equation}
 From Eq.~(\ref{eq:sh_eq_f_b}) and Eq.~(\ref{eq:factorized}),
we get
\begin{equation}
i(q\bar z+\bar z^{-1})
\frac
{\prod_{m=1}^{Q-1}(q\bar z-\bar z_m)}
{\prod_{m=1}^{Q-1}(\bar z-\bar z_m)}
- -i(q^{-1}\bar z+\bar z^{-1})
\frac
{\prod_{m=1}^{Q-1}(q^{-1}\bar z-\bar z_m)}
{\prod_{m=1}^{Q-1}(\bar z-\bar z_m)}
 =  E.
\label{eq:BA_derive}
\end{equation}
Comparing both sides and use the 
pole-free condition, then  one obtains the Bethe Ansatz equations
\begin{eqnarray}
\frac
{\bar z_l^2+q}
{q\bar z_l^2+1}
&=&
 -\prod_{m=1}^{Q-1}
\frac
{q \bar z_l - \bar z_m }
{\bar z_l - q \bar z_m }
,\ \ l=1,\cdots, Q-1
\label{eq:BA} \\
E &=& -i(q-q^{-1}) \sum_{m=1}^{Q-1}\bar z_m.
\label{eq:BA_e}
\end{eqnarray}
 From this equation, one can easily see that the 
energy $E$ is independent of $p_-$. Note that the functional 
solution corresponds to  $\Psi'_n$ is $\Psi(z^{-1})$,
which is not a polynomial and has poles.

\section{Explicit Solutions of the Bethe Ansatz equations for
$E=0$}
\label{sec:exp}
Let us consider from Eq.~(\ref{eq:BA_derive}). 
 For the zero energy $E=0$, 
Wiegmann and Zabrodin showed that $\Psi(z)$ is given by the
so-called continuous q-ultraspherical polynomial \cite{gr}
as $\Psi(z)= \frac{(q^2;q^2)_n}{(q;q^2)_n}
(-iz)^n P_n (-iz)$, where
$P_n (z) = \sum_{k=0}^n\frac{(q;q^2)_k(q;q^2)_{n-k}}
{(q^2;q^2)_k(q^2;q^2)_{n-k}}z^{n-2k}$, $n=(Q-1)/2$
and $(a;q)_k=\prod_{m=0}^{k-1}(1-aq^m)$.
\cite{another}
The q-ultraspherical polynomials have several interesting
properties. However, we treat the function $\Psi(z)$ directly. 

In order to understand the properties of the wavefunctions at
$E=0$, the center of the spectrum, we solve the Bethe Ansatz equation
(\ref{eq:BA}) explicitly. 
 For $E=0$, Eq.~(\ref{eq:BA_derive}) is written as
 and rewrite it as
\begin{equation}
q(z^2+q^{-1}) \Psi(qz)=q^{-1}(z^2+q)\Psi(q^{-1}z).
   \label{eq:dif}
\end{equation}
First put $z=iq^{+1/2}$, 
then one obtains $q(-q+q^{-1})\Psi(iq^{3/2})=0$ and
$\Psi(iq^{3/2})=0$.
So $i q^{ 3/2}$ is one of the roots.
Next put $z=iq^{+5/2}$, 
then one obtains $q(-q^5+q^{-1})\Psi(iq^{7/2})\propto
\Psi(iq^{3/2})=0$. Thus $i q^{ 7/2}$ is also a root. 
We can repeat this procedure and get a series of the roots as
$iq^{2m-1/2}$ where  $m=1,\cdots,(Q-1)/2$.
The restriction of $m$ arises due to the fact
that  a pre-factor of the $\Psi$
vanishes at the last step.

Similarly, put  $z=iq^{-1/2}$,
one obtains another sequence of the roots as
$iq^{-2m+1/2}$ where  $ m=1,\cdots,(Q-1)/2 $.
Comparing the number of the independent roots with the 
degree of the polynomial, $Q-1$, we have obtained a complete set
of roots as
\begin{equation}
\bar z_m  = 
\left\{
\begin{array}{l}
    i q^{ 2m-1/2} = i e^{i 2\pi \phi(m-1/4)}\\
    i q^{-2m+1/2} = i e^{i 2\pi \phi(m-1/4)}\\
\end{array} 
\right. 
\ m=1,\cdots,(Q-1)/2.
  \label{eq:allroots}
\end{equation}
They are all on the unit circle. 
Thus let us write them
as $z_m(p)=e^{i \theta_m(p)}$ and 
$\bar z_m=e^{i\bar \theta_m}$ and consider the distribution of
$\theta_m$ by 
$\rho(\theta)=\lim_{Q\rightarrow\infty}Q\Delta\theta$,
where $\Delta \theta$ is a difference between adjacent 
$\theta_m$.
Since $\bar z_m=z_m e^{i p}$,
$p$ dependence of the roots are simple shift of 
$\theta_m(p)=\bar \theta_m+p$.

 The restriction of $m$ ($ 1 \le m\le (Q-1)/2$) causes
a nontrivial distribution of the roots. 
The phase factor $\theta_m$ corresponds to a quasi-momentum in
the Bethe-Ansatz-solvable models in which $Q$
corresponds to the system size. 
 In the following, we consider $Q\to \infty$ limit which
corresponds to the thermodynamic limit.

Consider first the case $P=1$.
The roots $\{\bar z_m\}$ distribute uniformly on the unit
circle except near $z=\pm i$.
The roots for $Q=89$ are shown in Fig.~\ref{f:unif}.
In the semi-classical limit $Q\rightarrow \infty$, that is,
$q\rightarrow 1$,
the distribution function
$\rho(\theta)=\lim_{Q\rightarrow\infty}Q\Delta\theta$ is
smooth (constant).
A continuous behavior of $\rho(\theta)$ is usually obtained
in the Bethe-Ansatz-solvable models  in which $\rho(\theta)$ is
determined by an integral equation.

When the flux is irrational, the situation is quite different.
First let us take the flux $\phi=1/\tau=(\sqrt{5}-1)/2$, where
$\tau$ is the golden mean $(\sqrt{5}+1)/2$.
 To reach this flux, we consider a
sequence of rational fluxes $\phi_k=P_k/Q_k$, where
$Q_k=F_{3k+1},
\ P_k=F_{3k}$ and $F_k$ is a Fibonacci number defined by
$F_{k+1}=F_{k}+F_{k-1},\ F_1=1,\ F_0=1$. 
In this sequenses, $P_k$ and $Q_k$ are all odd. 
The two
types of roots in Eq.~(\ref{eq:allroots}) are nested and 
$\Delta\theta$ has a complex distribution. 
To gain an insight into the distribution of roots,
it is helpful to consider the pseudo roots which are defined
also by Eq.~(\ref{eq:allroots}) but with the range of $m$
modified to $m=-(Q-1)/2,\cdots,-1, 0$. 
In Figs.~\ref{f:gold}
we show the distributions of the roots (black points) and
pseudo roots (gray points) for several $\phi_k,(\ k=1,2,3,4,5,6)$.
Here the radius of the unit circle has been scaled so as
to show all the cases at once. These figures clearly show that
there is a branching rule for the true roots (denoted by $A$)
and pseudo roots (by $B$) as follows
\begin{eqnarray}
A^3 &\rightarrow & A^3B^2A^3B^2A^3 \nonumber \\
A^2 &\rightarrow & A^3B^2A^3 \nonumber \\
B^3 &\rightarrow & B^3A^2B^3A^2B^3 \nonumber \\
B^2 &\rightarrow & B^3A^2B^3.
  \label{eq:rule_gold}
\end{eqnarray}
The initial condition is $B^3A^2B^3A^2$ (cyclic).
In the $k$-th stage of the sequence,
the number of clusters of the true roots
$A^3$ and $A^2$ are $Q_{k-1}-1$ and
$P_{k-1}+1$ respectively. This branching rule gives
rise to a self-similar structure for the
distribution $\rho(\theta)$ in the limit $k \to\infty$.
To characterize the distribution, let us define the
generation of  roots. According to the branching
rule (\ref{eq:rule_gold}), each true (pseudo) root branches
into a cluster of 3 new true (pseudo) roots,
each of which in a sense
has a parent. At the same time, between these
clusters of new pseudo (true) roots, there is a pair of
new-born true (pseudo) roots which have no parent.
We assign the generation number to a root so that it
is $1$ when the root does not have a parent,
otherwise it is one plus the generation number of its parent.
Let us denote the number of true (pseudo) roots
in the $k$-th stage with generation $g$
by $n_A(g,k)\  (n_B(g,k))\ (g=1,\cdots,k)$.
(Then in the special case $P=1$,
$n_A(g,k)=n_B(g,k)=\delta_{kg}$.)
In the present case, there is a recursion formula
by the branching rule as
$n_A(g,k)=3 n_A(g-1,k-1), \ g=2, \cdots, k ,\ \
n_A(1,k)= 2 (P_{k-1} +1)$.
Thus $n_A(g,k)=2\cdot 3^{g-1} (P_{k-g}+1)$ and
$n_B(g,k)=2\cdot 3^{g-1} (P_{k-g}-1)$.

Next let us consider the flux
$\phi=1/\sigma=\sqrt{2}-1 $, where
$\sigma$ is the so-called silver mean $\sqrt{2}+1$ . 
To realize this flux in the large $Q$ limit, we
consider a sequence of rational fluxes $\phi_k=P_k/Q_k$, where
$Q_k=G_{k+1},
\ P_k=G_{k}$ and $G_k$ is defined by
$G_{k+1}=2 G_{k}+G_{k-1},\ G_2=1,\ G_1=1$. 
In this case,  we can apply
our Bethe Ansatz technique to each step of the sequence
since all $Q_k$ and $P_k$ are odd.  
 In Figs.~\ref{f:silver}
we show the distributions of the roots (black points) and
pseudo roots (gray points) for several 
$\phi_k,(\ k=1,2,3,4,5,6)$.
The radius of the unit circle are also scaled as before. 
One can observe  a clear branching rule branching. It is a
little different from the golden mean case and is given by
\begin{eqnarray}
  \label{eq:rule_silver}
A &\rightarrow  A^3 &\rightarrow  BA^2BA^2B  \nonumber \\
A^2 &\rightarrow  A^4 &\rightarrow  BA^2BA^2BA^2B \nonumber \\
B &\rightarrow  B^3 &\rightarrow  AB^2AB^2A  \nonumber \\
B^2 &\rightarrow  B^4 &\rightarrow  AB^2AB^2AB^2A 
\end{eqnarray}
The initial condition is $AB^2AB^2$ (cyclic). From this
branching rule, 
 we have a self similar behavior at every two stages.  
At the odd stage, all the roots and the pseudo roots appear as
$A$, $A^2$, $B$ and $B^2$ and at the even stage, they appear as
$A^3$, $A^4$, $B^3$, and $B^4$. 
For both cases, the distribution function of the roots  
has  self similarity and reflects the self similar
structure of the original problem. 

The above considerations clearly exemplify the difference
of the distributions between the semi-classical
limit and the incommensurate cases. The distribution
of the roots has a self-similar structure and
the function $\rho(\theta)$ is nowhere differentiable
in the incommensurate limit, while that for the
semi-classical limit is smooth. 
This is characteristic to the incommensurate case.

Another way to characterize the distribution  is
to map to the dual (reciprocal) space.
This can be done for arbitrary $P$ and $Q$ analytically.
We lift the $\theta_m$ to the real axis periodically.
On the real axis, the true and pseudo roots occupy
 lattice  points $\{j/2Q\ |\ j {\rm : integer} \}$
 with spacing $1/2Q$.
Thus we define the Fourier transform of the so-called defining
function 
by 
\begin{equation}
  S_Q(k)=\sum_{j=-\infty}^\infty e^{i k j} \tilde S_Q(j).
\label{eq:s_of_p}
\end{equation}
where $\tilde S_Q(j)=1$ if there is a true root at $j/2Q$,
otherwise $\tilde S_Q(j)=0$.
As shown in  Appendix A,
\begin{equation}
 S_Q(k)=\frac{\pi}{Q}
\sum_{r=0}^{Q-1}s_Q^r\,{\delta(k-k_r)}, \ (0\leq k<2 \pi ),
\label{eq:rootfrt}
\end{equation}
where 
\begin{equation}
s_Q^r=
\left\{
\begin{array}{ll}
(Q-1) & r=0 \\
\frac {(-)^{r+1}} { \cos(P k_r /2 )} & r=1,\cdots Q-1,
\end{array} 
\right. 
\label{eq:structure_factor}
\end{equation}
and
\begin{equation}
k_r=2\pi \frac r Q,
  \label{eq:def_mom}
\end{equation}
which converges to a usual continuous momentum $k\in [0,2\pi]$ 
in the $Q\to \infty $ limit.
In this limit, we have 
\begin{equation}
  \label{eq:sq}
   |S_Q(k)|^2=\frac 1 { 4 \cos(P k /2 )} , \ \ \  (0< k < 2\pi)
\end{equation}
In the semi-classical limit $P=1$ and $Q\to \infty$,
$|S_Q(k)|^2$  is well defined and behaves smoothly.
In the incommensurate limit, $|S_Q(k)|^2$ is a singular
function and is not even differentiable
 hence $P$ is infinite.
Also it can be shown that the original defining function is
given by
\begin{equation}
\tilde S_Q(j) = \frac 1 {2Q} \sum_{n=1}^{Q-1}
[ 1-(-)^n \frac {\cos(2\pi j n/Q)}{\cos(\pi P n/Q)}].
  \label{eq:def_function}
\end{equation}

\section{Explicit Wave functions by the Bethe Ansatz solutions}
\label{sec:exp_wfn}
Let us consider  wavefunctions which are obained from the
explicit solutions of the Bethe Ansatz equations. 
The wave function at site $j$  is given by $\Psi_j=\Psi_(q^j)$
and it is written in a compact factorized form as
 \begin{eqnarray}
\Psi_j&=& \prod_{m=1}^{(Q-1)/2} (q^j-q^{i(2m-1/)})(q^j-q^{-i(2m-1/)}) 
\nonumber \\
&=&  (-q)^{-j}(i q^{-j+3/2};q^2)_{(Q-1)/2}
(iq^{-j-3/2};q^{-2})_{(Q-1)/2},\\
&=& \sum_{\nu=0}^{(Q-1)/2} \sum_{\nu'=0}^{(Q-1)/2}
\left[\begin{array}{c}
        (Q-1)/2\\
          \nu
      \end{array}
\right]_q
\left[\begin{array}{c}
        (Q-1)/2\\
          \nu'
      \end{array}
\right]_q
(-iq^{-j})^{\nu+\nu'}
q^{Q(\nu-\nu')/2},
   \label{eq:wfna}
\end{eqnarray}
where we have used a $q$-binomial theorem
\begin{eqnarray}
  \prod_{n=1}^{N-1}(1+q^{N-1-2n}z)&=&\sum_{\nu=0}^N 
 \left[\begin{array}{c}
        N \\
          \nu
      \end{array}
\right]_q z^\nu ,\\
  \label{eq:binomial}
 \left[\begin{array}{c}
        \mu \\
        \nu
      \end{array}
\right]_q &=& \frac {[\mu]_q!}{[\nu]_q![\mu-\nu]_q!}
\label{eq:qconb}
\end{eqnarray}
It is convenient to shift the site, $j=\bar j+J_1$, by
an amount $J_1$
where $J_1$ is determined by 
$PJ_1=(Q-P)/2 \  ({\rm mod}\ 2Q\ )$. 
Then $\Psi_{\bar j}=0$ at
$\bar j=2m, -2m+1$ ($m=1,\cdots,(Q-1)/2$), and the wavefunction
is nonzero only at $\bar j=1,3,\cdots,Q$ and
$Q+1,Q+3,\cdots 2Q$ as shown in  Appendix B.
Thus one has
\begin{eqnarray}
|\Psi_{\bar 1}|^2 =Q, \\
|\Psi_{\bar j}|^2 =Q 
 \frac { [\bar j-2]_q!!}{ [\bar j-1]_q!!},
  \label{eq:amp}
\end{eqnarray}
for $\bar j=3,5,\cdots,Q$ and 
$|\Psi_{\bar j+Q}|^2=|\Psi_{\bar j}|^2$
where $[j]_q!!= [j]_q[j-2]_q\cdots [2]_q, \mbox{or} [1]_q$. 

When $P=1$, we can proceed further analytically.
As shown in the Appendix B, 
we get a compact form for the wavefunction
in the semi-classical limit, ( $P=1$ and $Q\to \infty)$,
\begin{equation}
|\psi (x)|^2 = \frac{2}{ \sin(\pi x)},\ \ \ (0<x<1),
  \label{eq:wfn}
\end{equation}
up to a constant factor, where
$\psi(x)=\Psi_{\bar j},\ x=(2j-1)/2Q$.
The squared amplitude of
the wavefunction is given by the
inverse chord distance. 
 The recursion relation
$|\Psi_{\bar j}|^2=|\Psi_{\bar j-2}|^2
\sin^2(\pi(j-2)/Q)/\sin^2(\pi(j-1)/Q)$
obtained from Eq.~(\ref{eq:amp})
has played a key role. \cite{ini}
 
 We can also calculate the  finite-size correction
 near the edge $x\approx 0$ or $x\approx 1$ for  $Q\to \infty$.
As shown in the appendix C, it is given by
\begin{eqnarray}
|\psi (x_{2l+1})|^2   &=& C(l)\frac {2}{ \sin\pi x_{2l+1}},\\
 C(l) &=&  \pi(l+\frac 1 4)\prod_{k=1}^l(1-\frac 1{2k})^2, \\
 &=&  \pi\frac{1+1/4l}{1+1/2l}\prod_{k=1}^l(1-\frac 1{4k^2}),
  \label{eq:sin}
\end{eqnarray}
where $C(0)=\pi/4=0.78539...,\ C(1)=5\pi/16=0.98174...,
\ C(2)=81\pi/256=0.99402...,\cdots$.
Since $\sin z=z\prod_{k=1}^\infty(1-\frac{z^2}{k^2\pi^2})$,
$\prod_{k=1}^\infty(1-1/4k^2)=\pi/2$.
So the finite-size correction factor $C(l)$ converges to unity
very rapidly and the Eq.~(\ref{eq:wfn}) is quite accurate even
for small $Q$. The norm of the wavefunction is $\log Q+{\rm
const.}$ and unnormalizable which is characteristic in a
critical wavefunctions. 
In Fig.~\ref{f:wfnsemi}, the amplitudes of the analytic
wavefunctions, normalized by the peak height, are shown for
several values of $Q$.

Next let us discuss the case with golden-mean flux.
We plot the analytic results  Eq.~(\ref{eq:amp})
in Fig.~\ref{f:wfninc} for a sequence of rational
fluxes converging to $1/\tau$. One can easily recognize
the self-similar behavior of the wave function.
Each peak branches into three peaks in the next stage.
Presumably these are the reflection of the
self-similar distribution of the roots, 
Eq.~(\ref{eq:rule_gold}).

We have performed a multifractal analysis\cite{hk} to
investigate the nature of the wavefunctions.
This is useful to distinguish critical wavefunctions from extended
wavefunctions.
The results for critical wavefnctions reveals
the multifractal properties. 
 Let us consider a $k$-th generation $Q=Q_k=F_{3k+1}$.
First define a probability measure
$p_j$ ($\sum_{j} p_j=1$ ) of the wave function as
\begin{eqnarray}
\label{eq:prob}
p_j&=&\frac 1 {{N_k}}  |\Psi_j |^2 \\ 
N_k&=& \sum_{j=1}^{Q_k} |\Psi_j |^2   .
\end{eqnarray}
 Next define a Lebesgue measure
$l_k$ of each site as 
\begin{equation}
\label{eq:leb}
l_k=\frac 1{Q_k}.
\end{equation}
From $p_j$ and $l_k$, the singularity of the probability
measure is represented by an exponent $\alpha_j$ as
\begin{equation}
\label{eq:alpha}
p_j=l_k^{\alpha_j }.
\end{equation}
The distribution $\Omega_k(\alpha)$ characterizes the wave
function, where $\Omega_k(\alpha)d\alpha$ is the number of
sites whose value of
$\alpha_j$ lies in the interval $[\alpha, \alpha+d\alpha]$. 
Since $Q_k$ increases exponentially as $k$ increases, so does
$\Omega_k$.
 Thus it is natural to introduce  the entropy function
$S(\alpha)$ to characterize the wave function as \cite{several_mk}
\begin{equation}
\label{eq:entropy}
S(\alpha)=\lim_{k\to\infty}  S_k(\alpha)
=\lim_{k\to\infty}\frac 1 k \log \Omega_k(\alpha),
\end{equation}
or
\begin{equation}
\label{eq:falpha}
f(\alpha)=\frac {S(\alpha) }\epsilon,
\end{equation}
where
\begin{equation}
\label{eq:epsilon}
\epsilon=\lim_{k\to\infty} \epsilon_k=-\lim_{k\to\infty}
 \frac 1 k 
\log l_k.
\end{equation}
Thus we have 
$
\Omega_k(\alpha)\approx l_k^{-f(\alpha)}
$.
To calculate  $f(\alpha)$ let us define   the
partition function
$Z_k(r)$ as
\begin{equation}
\label{eq:part}
Z_k(r)=\sum_{j=1}^{Q_k}p_j^r=\sum_{j=1}^{Q_k}
l_k^{-r k \alpha_j \epsilon_k} .
\end{equation}
One obtain $f(\alpha)=\lim_{k\to \infty}f_k(\alpha) $ from  
\begin{eqnarray}
\label{eq:calcf}
G_k(r)&=&\frac 1 k \log Z_k(r), \\
\alpha & =& -\frac 1 {\epsilon_k} \frac d {dr} G_k(r),\\
f_k(\alpha)&=&\frac 1 {\epsilon_k} G_k(r) + r \alpha. 
\end{eqnarray}
When the wave function is extended, $p_j\approx l_k$ and
$f(\alpha=1)$ at $\alpha=1 $.
On the other hand, when it is localized, $f(\alpha)$ consists
of two points at $\alpha=0$ and $\alpha=\infty$. 
When $f(\alpha)$ is a continuous function, the wave
function is critical and  has a singular probability measure
which is a characteristic of the multifractality.
We have performed the above multifractal analysis. 
Here we stress the importance of
the finite size effects. 
In order to have a reliable $f(\alpha)$,
one must perform extraporations of the finite size data.
($f(\alpha)$ obtained from a finite system 
is different from the true $f(\alpha)$ \cite{hk,fkt}.)
We have done such calculations. 
  The results
is shown in Fig.~\ref{f:mf}
for a golden mean flux.
It gives a smooth $f(\alpha)$.
This clearly shows that this wavefunction is multifractal and
critical.
 We note the striking resemblance of these
wavefunctions to that of the $1d$ quasicrystal Fibonacci
lattice at the center of the spectrum.\cite{kb}
The latter was obtained exactly by a different technique and
$f(\alpha)$ is obtained analytically.\cite{fkt}

\section{Numerical Solutions of the Bethe Ansatz equations for 
$E\neq0$}
\label{sec:num}

For $E\neq 0$ cases, we have not being able to obtain analytical
results.  
Thus we  investigate them.
However, the Bethe Ansatz equations are high degree
algebraic equations of many variables and it is extremely
difficult to handle directly even numerically. 
Thus we  use the information of the quantum group to reduce
the difficulty. 
Instead solving the Bethe Ansatz equations,
Eq.~(\ref{eq:BA}), we construct a polynomial
$\Psi^j(z)$ for a mid band energy $E^j$ when $\Psi^j$ is written 
as
When one write the $\Psi(z)$ as 
\begin{equation}
  \label{eq:psi_z}
  \Psi^j(z)=\sum_{m=0}^{Q-1} v_m^j z^m,
\end{equation}
$[v_0^j,\cdots,v_{Q-1}^j]^T$ is given by the $j$-th eigen
vector of  the $Q\times Q$ tri-diagonal matrix
$H_{MB}^{tri}$.        
The roots of the Bethe Ansatz equations
are given by the roots of this  one variable equation.
 Since the $H_{MB}^{tri}$ is real symmetric, 
$  \Psi^j(z)$ is a real polynomial. 
Thus the roots can be  obtained by the  traditional
numerical  technique.

Let us consider the  cases $P=1$ first. 
 We have calculated the roots of the Bethe Ansatz for
substantially many different $Q$.
All roots are on the unit circle.  
It seems  that 
all roots of the Bethe Ansatz equations are on the unit circle
when
$\phi=1/Q$ with odd $Q$. 
 In Fig.~\ref{f:several_semi}, we  present results for several
cases with $Q=61$. 
In Fig.~\ref{f:all_semi}, all root for $Q=41$ are shown. 
At the highest energy band, the roots are on the  right
semi-circle of the unit circle and it is almost uniformly
arranged (not exactly for finite $Q$). 
At the second energy band, one root appears in the left
semi-circle (at $z=-1$).
At the third one, one more roots appear in the left. 
In this way, one of the roots are shifted from right
semi-circle to the left semi-circle as the band energy
decreases. This behavior is clearly observed in 
Fig.~\ref{f:all_semi}.
We have also calculated the distribution functions
which are plotted in  Fig. ~\ref{f:dist_semi}.
For a finite $Q$ case, we define an approximate distribution
function $\rho_\epsilon^{Q}(\theta)$ as
\begin{eqnarray}
\label{eq:dist}
\rho_\epsilon^{Q}(\theta)&=&
\sum_{m=1}^{Q-1}
\delta_\epsilon(\theta-\theta_m),\\
\delta_\epsilon(x)&=&-\frac 1 \pi 
\Im(\frac 1 {2 \sin(\frac { x}{2} )-i\epsilon}),
\end{eqnarray}
where $\epsilon$ is a width of the approximate delta functions
and we take as $\epsilon=2/Q$. The distance is measured by the
chord distance.
In Fig. ~(\ref{f:dist_semi}),
we plot the results for several cases. 
 From the results for the  finite $Q$, we can speculate a
behavior of the distributions in the $Q\to
\infty$ limit.  
It is likely to have singularities for $E\neq 0$ states
 at $z=\pm i$ which was
proved for $E=0$ state.


\section{acknowledgment}
This work was supported in part by Grant-in-Aid
from the Ministry of Education, Science and Culture 
of Japan. Y.-S.W. thanks both this grant and the Institute 
for Solid State Physics, University of Tokyo for support 
and warm hospitality during his visit, when this collaboration
began. He also thanks R. Musto and H.C. Fu for helpful
discussions, Y.H. and M. K. are grateful to P. Wiegmann for 
valuable discussions.

\appendix
\section{Derivation of the structure factor of the Bethe Ansatz roots}
\label{sec:append1}
Here we  derive Eq.~(\ref{eq:rootfrt})-Eq.~(\ref{eq:def_mom}).
Using Eq.~(\ref{eq:allroots}), one obtains 

\begin{eqnarray}
 S_Q(k)& =&\sum_{j=-\infty}^\infty e^{i k j} \tilde S_Q(j) \nonumber \\
 & =& \sum_{l=-\infty}^{\infty}\sum_{m=1}^{Q-1}
[ e^{i k \{ 2Ql+2Pm+\frac 1 2 (Q-1P) \}} 
+
e^{i k \{ 2Ql -2 P m + (Q+P)/2 \}}  ]  \nonumber \\
&=&\sum_{l=-\infty}^{\infty} 
e^{i 2 k Ql} \left\{
\frac{1-e^{i  P k (Q-1)}}{1-e^{i2Pk}}e^{i(Q+3P)k/2}
+
\frac{1-e^{-i  P k (Q-1)}} {1-e^{-i2Pk}} e^{i(Q-3P)k/2} \right\} 
\nonumber  \\
& = & 
\sum_{r^\prime =0}^{2Q-1}
\frac {2 \pi} {2Q} \delta (k-\frac { 2\pi r^\prime}{2Q})
\frac
{\sin P(Q-1)k/2}
{\sin Pk}
\ e^{i Qk/2} \ 2\cos\frac 1 2 PQk.
\end{eqnarray}
The structure factor is a sum of the delta function but 
the amplitude of
the peak
has a nontrivial feature. 
At $k=\frac { 2\pi r^\prime} {2Q}$, we have
\begin{equation}
\frac
{\sin P(Q-1)k/2}
{\sin Pk}
\  e^{i Qk/2} \ 2\cos\frac 1 2 PQk
  = \left\{
  \begin{array}{l c}
Q-1 & \ \  r^\prime = 0\ \  \\
\frac{(-)^r} { \cos \pi \frac P Q r} & \ \  r^\prime =2 r \neq 0 \\
0 & \ \ otherwise
  \end{array}
\right.
\end{equation}
This leads to Eq.~(\ref{eq:rootfrt})-Eq.~(\ref{eq:def_mom}). 
In Eq.~(\ref{eq:structure_factor}),
 $\cos P k/2$ is smooth function of $k\in (-\pi,\pi]$
when P is finite. It, however,
 behaves wildly when $P\to \infty$.
This case 
corresponds
to an irrational flux.

\section{Explicit wave functions in a closed form}
\label{sec:append2}
 From the explicit solution of the Bethe Ansatz equation,
the  wave function at site $j$ is written as
\begin{eqnarray}
  \psi_j & = & \prod_{m=1}^{(Q-1)/2}(q^j-iq^{2m-1/2})
(q^j-iq^{-2m+1/2}) \nonumber \\
         & = & q^{(Q-1)j}
\prod_{m=1}^{(Q-1)/2}(1-iq^{2m-1/2-j})(q^j-iq^{-2m+1/2-j})  \nonumber  \\
         & = & (-q)^{-j}
(iq^{3/2-j};q^2)_{(Q-1)/2}(iq^{-3/2-j};q^{-2})_{(Q-1)/2}.
\end{eqnarray}
The two sequences of the roots are 
obtained explicitly and we have
\begin{eqnarray}
  \psi_j 
         =  q^{(Q-1)j}  \prod_{m=1}^{(Q-1)/2} &&
(1-\exp[{i 2\pi\frac 1 {2Q} \{ P (2m-j) + (Q-P)/2 \}} ] ) 
 \nonumber \\
&\times & (1-\exp[{i 2\pi\frac 1 {2Q} \{ P (-2m-j) + (Q+P)/2 \}} ])
\end{eqnarray}
Let us define $J_1$ by 
\begin{equation}
  PJ_1\equiv \frac 1 2 (Q-P) \pmod{2Q},
\end{equation}
then $P(J_1+1)\equiv \frac 1 2 (Q+P) \pmod{2Q}$ and 
$
  \psi_j=0, \ {\rm for}\  j \equiv  2m+J_1, -2m+J_1+1
$.
$
 (m  = 1,\cdots (Q-1)/2).
$
Thus we have 
\begin{equation}
  \psi_{\bar j}  =  0, \ {\rm for \ } \ \  \bar j = 2m,
 \ {\rm or } -2m+1
(  m  = 1,\cdots (Q-1)/2)
\end{equation}
where $ \bar j =j-J_1 $.

Now let us consider the amplitude of the wave function.
Apparently period of $|\psi_{\bar j}|$ is $2Q$.
However, it is not the smallest
period. 
Actually the smallest period is $Q$ 
because  $\{z_m\}=\{z_m^*\}$.
Thus define $ l$ by $\bar j = 2l-1$ and
consider $\psi(l)=\psi_{\bar j= 2l-1}\ (j=J_1+2l-1$, 
$l=1,2,\cdots,(Q+1)/2$).
Then the squared amplitude is 
\begin{eqnarray}
|\psi(l)|^2
         =   \prod_{m=1}^{(Q-1)/2} && 
|1-\exp[{i 2\pi\{\frac P {2Q} (2m-1)-\frac P Q (l-1)\} } |^2 
\nonumber \\
&\times & |1-\exp[{i 2\pi\{\frac P {2Q}  
(-2m)-\frac P Q (l-1)\} }]|^2 \nonumber  \\
         =   \prod_{m=1}^{(Q-1)/2}  
&& 
(1-q^{-2(l-1)}q^{2m-1})(1-q^{2(l-1)}q^{-(2m-1)}) \nonumber  \\
&\times&(1-q^{-2(l-1)}q^{-2m})(1-q^{ 2(l-1)}q^{2m}).  \\
  \label{eq:general_l}
\end{eqnarray}

Let us first consider $|\psi(l)|$ for $l=1$. 
Using  
\begin{eqnarray}
S_1&=& \{2m-1\ | m=1,\cdots,(Q-1)/2\} = \{1,3,\cdots,Q-4,Q-2\} 
\nonumber \\
S_2&=&\{-(2m-1)\ | m=1,\cdots,(Q-1)/2\} = 
\{-1,-3,\cdots,-(Q-4),-(Q-2)\} 
\nonumber \\
& \equiv  &
\{Q+2,Q+4\cdots,2Q-3,2Q-1\} \pmod{2Q}\nonumber  \\
S_3&=&\{-2m\ | m=1,\cdots,(Q-1)/2\} = \{-2,-4,\cdots,-(Q-3),-(Q-1)\} 
\nonumber  \\
&\equiv  &
\{Q+1,Q+3\cdots,2Q-4,2Q-2\} \pmod{2Q} \nonumber  \\
S_4&=&\{2m\ | m=1,\cdots,(Q-1)/2\} = \{2,4,\cdots,(Q-3),(Q-1)\}
,\nonumber
\end{eqnarray}
we have 
\begin{equation}
S=S_1\cup S_2\cup S_3\cup S_4\equiv \{m\ | m=1,\cdots,2Q\}\setminus \{0,Q\} .
\pmod{2Q}
  \label{eq:set2}
\end{equation}
Since $P$ and $2Q$ are mutually prime,
$S$ is invariant under a multiplication by $P$ $\pmod{2Q}$.
This leads to
\begin{eqnarray}
   \prod_{m=1}^{(Q-1)/2}  
(1-q^{2m-1})(1-q^{-(2m-1)})(1-q^{-2m})(1-q^{2m})
&=&\prod_{
m\in\{1,\cdots,2Q\}\setminus \{0,Q\}
}
  (z-e^{i 2\pi \frac{m}{2Q}}) \nonumber  \\
&=&\frac {z^{2Q}-1}  {(z-1)(z+1)} .
  \label{eq:sum1}
\end{eqnarray}
In a limit $z\to 1$, we have
\begin{equation}
|\psi(1)|^2      =   Q,
  \label{eq:l_1=Q}
\end{equation}
which is independent of $P$.

In order to obtain the other amplitude, we use
a resursion relation which is obtained from 
Eq.~(\ref{eq:general_l}):
\begin{eqnarray}
|\psi(l+1)|^2
&=& |\psi(l)|^2
\frac {1-q^{-2(l-1)-1}}{1-q^{-2(l-1)+Q-2}}
\frac {1-q^{2(l-1)+1}}{1-q^{2(l-1)-Q+2}}\nonumber \\
&& \ \ \ \ \times 
\frac {1-q^{-2(l-1)-(Q+1)}}{1-q^{-2(l-1)-2}}
\frac {1-q^{2(l-1)+(Q+1)}}{1-q^{2(l-1)+2}}\nonumber \\
&=&
 |\psi(l)|^2 \left[
\frac 
{\sin \pi \frac P Q (2l-1)}
{\sin \pi \frac P Q 2l}
\right]^2.
  \label{eq:generic}
\end{eqnarray}
 From this we have 
\begin{equation}
\psi(l)=\pm Q \frac { [2l-1]_q!!}{ [2l]_q!!},
  \label{eq:wfnexp}
\end{equation}
where $[2n]_q!!= [2]_q[4]_q\cdots [2n]_q$ and 
$[2n-1]_q!!= [1]_q[3]_q\cdots [2n-1]_q$. 

Now fix $P=1$ and take a large $Q$ limit.
The continuum coordinate $x_l$ and the square amplitudes
 are defined by
\begin{eqnarray}
x_l&=&(l-\frac 3 4 ) \Delta x,\  \ ( l=1,\cdots,(Q+1)/2) \\
\Delta x & = & \frac 2 Q \\
n(x_l)&\equiv&|\psi(l)|^2.
  \label{def_x}
\end{eqnarray}
In the large $Q$ limit, $x$ is in an interval 
$(0,1)$.
Taking Eq.~(\ref{eq:generic}) upto first order in $\Delta x$
and we get 
\begin{equation}
\frac d {dx} \log n(x) =-\frac d {dx} \sin \pi x.
  \label{eq:diff_eq}
\end{equation}
Thus 
\begin{equation}
n(x)=\frac C{\sin \pi x},
  \label{eq:n}
\end{equation}
where $C$ is a some constant.
The coefficient $C$  is determined by 
considering a case  $l={\cal O}(1)<<Q $ where
\begin{eqnarray}
n(x_k)&=&Q \prod_{j=1}^{k-1}
\left[ \frac 
{\sin 2\pi \frac P Q (j-\frac 1 2)}
{\sin 2\pi \frac P Q j}\right ] \nonumber \\
&=& Q\prod_{j=1}^{k-1}
\left(
\frac {j-\frac 1 2}{j}
\right)^2 
  \label{constant}
\end{eqnarray}

\section{Derivation of the finite size correction for the large $Q$}
\label{sec:append3}

We derive the finite size correction near the edges
of the wave function for $P=1$ and $E=0$. 
Write the amplitude as 
$|\psi (x_{l})|^2  
=C(l)\frac {2}{ \sin\pi x_{2l+1}}$, 
where $x_{2l+1}=(l+1/4)2/Q$.
The finite size correction $C(l)$  near the edge
is given by 
\begin{eqnarray}
\label{eq:cor}
C(l)&=&{\frac  1 2} |\psi (x_{l})|^2 \sin\pi x_l \nonumber \\
    &=&{\frac  1 2} Q
 (\frac { [2l-1]_q!!}{ [2l]_q!!} )^2\sin{\frac {2\pi} Q} 
(l+{\frac 1 4}) \nonumber \\ 
& \rightarrow   & Q \prod _{k=1}^l({\frac {2k-1}{2k}})^2 {\frac \pi Q}
(2l+\frac 1 2)
\  \ \ \ \ \ \ ({\frac l Q} \rightarrow 0) \nonumber \\
& = & \frac \pi 2 (2l+{\frac 1 2})\prod_{k=1}^l {\frac {2k-1} {2k+1} }
{ \frac {(2k-1)(2k+1)} {4k^2 }} \nonumber \\
& = & \frac \pi 2{\frac{2l+1/2} {2l+1}} \prod_{k=1}^l(1-{\frac 1 {4k^2}}),
\end{eqnarray}
where $l$ finite but is large but $Q\to \infty$.

\begin{figure}
\caption{
The energy spectrum of the Bethe Ansatz equations 
in the complex plane for (a) $Q=23$
and $P=1,3,5,\cdots,Q-3,Q-1$ ($+$) and (b)  $Q=401$
and$P=1,3,5,\cdots,Q-3,Q-1$ ($\cdot$ ) . 
 \label{f:ene}}
\end{figure}

\begin{figure}
\caption{The roots of the Bethe ansatz equation 
for the  case
$P=1$ and $Q=89$ in the complex plane.
 \label{f:unif}}
\end{figure}

\begin{figure}
\caption{The roots and pseudo roots 
of the Bethe ansatz equation for the
rational fluxes which converges to $1/\tau$,
$\phi_k=P_k/Q_k=3/5,$$ 13/21,$$ 55/89,$$ 233/377,$
$ 987/1597,$$ 4181/6765$
(a) in the whole complex plane and (b) an enlarged figure.
 In each case,
the roots and the pseudo roots 
are always on the unit circle. 
The radii for $\phi_k$ are scalled to
show the branching rule.
 \label{f:gold}}
\end{figure}

\begin{figure}
\caption{The roots of the Bethe ansatz equation for the
rational fluxes which converges to $1/\sigma$,
$\phi_k=P_k/Q_k=3/7,$$ 7/17,$$ 17/41,$$ 41/99,$$ 99/239,$$
239/577$, and $577/1393$ (a) in the whole complex plane and (b)
an enlarged figure. The radii are scaled.
 \label{f:silver}}
\end{figure}

\begin{figure}
\caption{Squared amplitudes for the wavefunctions for $P=1$ and
$Q=5,21,89,377$ 
.
 The wavefunctions are normalized by the peak
heights.
 \label{f:wfnsemi}}
\end{figure}

\begin{figure}
\caption{Amplitude of the wavefunctions: for the ratios of
successive Fibonacci numbers:
(a) whole region ($\phi=3/5,$ $ 13/21,$ $ 55/89 $ )
and
(b) enlarged plot in the region near 0.89.
($\phi=P_k/Q_k=3/5,$ $ 13/21,$ $ 55/89,$ $ 233/377,$ $ 987/1597,$
$ 4181/6765$)
The wavefunction with larger value of $Q$ is shaded darker.
The wavefunctions are normalized by the peak height.
 \label{f:wfninc}}
\end{figure}

\begin{figure}
\caption{ $f(\alpha)$ for the wavefunction at 
$E=0$ for  the golden mean
case.
 \label{f:mf}}
\end{figure}

\begin{figure}
\caption{The roots of the Bethe Ansatz equations for $E\neq 0$
states 
in the complex  plane
for  $P=1$ and $ Q=61$. (a) The highest energy
state, (b) the second higest energy state,
 (c) the third highest energy state,
and (d) the 30-th energy state which is just one above the
center ($E=0$) state. 
 \label{f:several_semi}}
\end{figure}

\begin{figure}
\caption{All sets of the roots of the Bethe Ansatz equations
in the complex plane
for $P=1$ and $ Q=41$.  All the roots are on the unit circle.
 We scale the radii to show them together. The radius of the
higher state scaled smaller. 
 \label{f:all_semi}}
\end{figure}

\begin{figure}
\caption{Approximated distribution function of the roots of
the Bethe Ansatz equations for $P=1$ and $ Q=41$. 
 (a) The highest energy
state, (b) the 3-rd energy state, (c) the 10-th energy state,
(d) the 15-th energy state, (e) the 18-th energy state,
and (f) the 21-th energy state which is the
center ($E=0$) state. 
 \label{f:dist_semi}}
\end{figure}

\begin{figure}
\caption{The roots of the Bethe Ansatz equations for $E\neq 0$
states for $P=55$ and $ Q=89$. (a) The highest energy
state, (b) the 22-nd energy state, (c) the 30-th energy state,
and (d) the 34-th energy state.
 \label{f:several_fib}}
\end{figure}


\begin{references}
\bibitem{another}
Our 1d difference equation ( in the diagonal gauge) at $E=0$
has another solution which is given by
$
\tilde \Psi_l=\tilde \Psi(q^l)$ where
$\tilde \Psi(z)=  \frac{(q^2;q^2)_n}{(q;q^2)_n}$
$ (-iz^{-1})^{n+1} P_n(iz^{-1})$
which has a pole at $z=0$.
This solution does not satisfy the pole free condition,
and separate consideration is needed.
However, since $\{z_m\}=\{z_m^*\}$ as shown later,
we have $|\Psi_l|=|\tilde\Psi_l|$. Thus we only consider
$\Psi_l$ from now on.
\end{references}
\end{document}